
\magnification=\magstephalf\nopagenumbers
\hsize=6.5 truein
\vsize=9.5 truein
\hfuzz=2pt\vfuzz=4pt
\pretolerance=5000
\tolerance=5000
\parskip=0pt plus 1pt
\parindent=16pt
\font\fourteenrm=cmr10 scaled \magstep2
\font\fourteeni=cmmi10 scaled \magstep2
\font\fourteenbf=cmbx10 scaled \magstep2
\font\fourteenit=cmti10 scaled \magstep2
\font\fourteensy=cmsy10 scaled \magstep2

\font\big=cmr10 scaled \magstep1
\font\large=cmr10 scaled \magstep2

\font\sixrm=cmr6

\font\eightrm=cmr8
\font\eighti=cmmi8
\font\eightbf=cmbx8
\font\eightit=cmti8

\font\eightsy=cmsy8
\font\sixrm=cmr6
\font\sixi=cmmi6
\font\sixsy=cmsy6

\def\tenpoint{\def\rm{\fam0\tenrm}%
 \textfont0=\tenrm \scriptfont0=\sevenrm
	   \scriptscriptfont0=\fiverm
 \textfont1=\teni \scriptfont1=\seveni
	   \scriptscriptfont1=\fivei
 \textfont2=\tensy \scriptfont2=\sevensy
	   \scriptscriptfont2=\fivesy
 \textfont3=\tenex  \scriptfont3=\tenex
	   \scriptscriptfont3=\tenex
 \textfont\itfam=\tenit \def\it{\fam\itfam\tenit}%
 \textfont\slfam=\tensl \def\sl{\fam\slfam\tensl}%
 \textfont\bffam=\tenbf \scriptfont\bffam=\sevenbf
	      \scriptscriptfont\bffam=\fivebf
	      \def\bf{\fam\bffam\tenbf}%
 \normalbaselineskip=20 truept
 \setbox\strutbox=\hbox{\vrule height14pt depth6pt width0pt}%
 \let\sc=\eightrm \normalbaselines\rm}
\def\eightpoint{\def\rm{\fam0\eightrm}%
 \textfont0=\eightrm \scriptfont0=\sixrm
	   \scriptscriptfont0=\fiverm
 \textfont1=\eighti \scriptfont1=\sixi
	   \scriptscriptfont1=\fivei
 \textfont2=\eightsy \scriptfont2=\sixsy
	   \scriptscriptfont2=\fivesy
 \textfont3=\tenex  \scriptfont3=\tenex
	   \scriptscriptfont3=\tenex
 \textfont\itfam=\eightit \def\it{\fam\itfam\eightit}%
 \textfont\bffam=\eightbf \def\bf{\fam\bffam\eightbf}%
 \normalbaselineskip=16 truept
 \setbox\strutbox=\hbox{\vrule height11pt depth5pt width0pt}}
\def\fourteenpoint{\def\rm{\fam0\fourteenrm}%
 \textfont0=\fourteenrm \scriptfont0=\tenrm
	   \scriptscriptfont0=\eightrm
 \textfont1=\fourteeni \scriptfont1=\teni
	   \scriptscriptfont1=\eighti
 \textfont2=\fourteensy \scriptfont2=\tensy
	   \scriptscriptfont2=\eightsy
 \textfont3=\tenex  \scriptfont3=\tenex
	   \scriptscriptfont3=\tenex
 \textfont\itfam=\fourteenit \def\it{\fam\itfam\fourteenit}%
 \textfont\bffam=\fourteenbf \scriptfont\bffam=\tenbf
	       \scriptscriptfont\bffam=\eightbf
	       \def\bf{\fam\bffam\fourteenbf}%
 \normalbaselineskip=24 truept
 \setbox\strutbox=\hbox{\vrule height17pt depth7pt width0pt}%
 \let\sc=\tenrm \normalbaselines\rm}

\def\today{\number\day\ \ifcase\month\or
 January\or February\or March\or April\or May\or June\or
 July\or August\or September\or October\or November\or December\fi
 \space \number\year}
\newcount\secno   
\newcount\subno   
\newcount\subsubno  
\newcount\appno   
\newcount\tableno  
\newcount\figureno  
\normalbaselineskip=15 truept
\baselineskip=15 truept

\def\title#1#2
  {\vglue0.5truein
  {\centerline{\large #1}\smallskip\centerline{\large #2}
  \vskip 0.5truein}}
\def\author#1{\centerline{\big #1}\medskip}
\def\address#1{\centerline{\it #1}\medskip}

\def\shorttitle#1
  {\vfill
  \noindent \rm Short title: {\sl #1}\par
  \medskip}
\def\pacs#1
  {\noindent \rm PACS number(s): #1\par
  \medskip}
\def\jnl#1
  {\noindent \rm Submitted to: {\sl #1}\par
  \medskip}
\def\date
  {\noindent{Date: \today}\medskip}
\def\keyword#1
  {\bigskip
  \noindent {\bf Keyword abstract: }\rm#1}
\def\beginabstract
  {\vskip 0.5truein\baselineskip=12pt
  \noindent{\bf Abstract}\smallskip\rm}
\def\endabstract
  {\bigskip}

\def\appendix#1
  {\vskip0pt plus.1\vsize\penalty-250
  \vskip0pt plus-.1\vsize\vskip24pt plus12pt minus6pt
  \subno=0 \eqnno=0
  \global\advance\appno by 1
  \noindent {\bf Appendix \the\appno. #1\par}
  \bigskip
  \noindent}
\def\subappendix#1
  {\vskip-\lastskip
  \vskip36pt plus12pt minus12pt
  \bigbreak
  \global\advance\subno by 1
  \noindent {\sl \the\appno.\the\subno. #1\par}
  \nobreak
  \medskip
  \noindent}
\def\ack
  {\vskip0pt plus.1\vsize\penalty-250
  \vskip0pt plus-.1\vsize\vskip18pt plus9pt minus6pt
  \bigbreak
  \noindent{\bf Acknowledgements\par}
  \nobreak
  \bigskip
  \noindent}

\def\tabcaption#1
  {\global\advance\tableno by 1
  \noindent {\bf Table \the\tableno.} \rm#1\par
  \bigskip}

\def\figcaption#1
  {\global\advance\figureno by 1
  \noindent {\bf Figure \the\figureno.} \rm#1\par
  \bigskip}
\def\references
   {\vskip0pt plus.1\vsize\penalty-250
  \vskip0pt plus-.1\vsize\vskip18pt plus9pt minus6pt
   \noindent{\bf References}\par
   \parindent=0pt
   \bigskip}
\def\refjl#1#2#3#4
  {\hangindent=16pt
  \hangafter=1
  \rm #1
  {\frenchspacing\sl #2
  \bf #3}
  #4\par}
\def\refbk#1#2#3
  {\hangindent=16pt
  \hangafter=1
  \rm #1
  {\frenchspacing\sl #2}
  #3\par}
\def\numrefjl#1#2#3#4#5
  {\parindent=40pt
  \hang
  \noindent
  \rm {\hbox to 30truept{\hss #1\quad}}#2
  {\frenchspacing\sl #3\/
  \bf #4}
  #5\par\parindent=16pt}
\def\numrefbk#1#2#3#4
  {\parindent=40pt
  \hang
  \noindent
  \rm {\hbox to 30truept{\hss #1\quad}}#2
  {\frenchspacing\sl #3\/}
  #4\par\parindent=16pt}

\def\frac#1#2{{\displaystyle{#1 \over #2}}}

\def\d{{\rm d}}

\def\i{\ifmmode{\rm i}\else\char"10\fi}
\def\case#1#2{{\textstyle{#1\over #2}}}

\def\etal{{\sl et al.\/}\ }
\catcode`\@=11
\def\ind{\hbox to 5pc{}}
\def\eq(#1){\hfill\llap{(#1)}}

\def\deqn#1{\displ@y\halign{\hbox to \displaywidth
  {$\@lign\displaystyle##\hfil$}\crcr #1\crcr}}
\def\indeqn#1{\displ@y\halign{\hbox to \displaywidth
  {$\ind\@lign\displaystyle##\hfil$}\crcr #1\crcr}}
\def\indalign#1{\displ@y \tabskip=0pt
 \halign to\displaywidth{\ind$\@lign\displaystyle{##}$\tabskip=0pt
  &$\@lign\displaystyle{{}##}$\hfill\tabskip=\centering
  &\llap{$\@lign##$}\tabskip=0pt\crcr
  #1\crcr}}
\catcode`\@=12


\font\sc=cmcsc10

\font\shell=msbm10

\def\N{{\hbox{\shell N}}}
\def\Na{{\hbox{\shell N}}}

\def\Z{{\hbox{\shell Z}}}

\def\tx{\textstyle}

\def\tfr#1#2{{\tx{#1\over#2}}}

\def\p{Pain\-lev\'e}

\def\bt{B\"ack\-lund transformation}
\def\bts{B\"ack\-lund transformations}

\def\erfc{\mathop{\rm erfc}\nolimits}

\newcount\exno
\newcount\secno  
\newcount\subno  
\newcount\subsubno  
\newcount\figno
\newcount\tableno

\def\section#1
  {\vskip0pt plus.1\vsize\penalty-250
  \vskip0pt plus-.1\vsize\vskip18pt plus9pt minus6pt
  \subno=0 \exno=0 \figno=0 
  \global\advance\secno by 1
  \noindent{\bf\the\secno. #1}
  \medskip}
\def\subsection#1
  {\vskip-\lastskip
  \exno=0 \caseno=0 \tableno=0 \subsubno=0
  \vskip15pt plus4pt minus4pt
  \bigbreak
  \global\advance\subno by 1
  \noindent{\sl \the\secno.\the\subno. #1}\smallskip}

\def\subsubsection#1
  {\vskip-\lastskip
  \exno=0 \caseno=0
   \vskip4pt plus2pt minus2pt
  \bigbreak \global\advance\subsubno by 1
  \noindent {\sl \the\secno.\the\subno.\the\subsubno\enskip #1. }}

\def\=#1{{\bf\bar{\mit#1}}}
\def\^#1{{\widehat{\mit#1}}}
\def\~#1{{\widetilde{\mit#1}}}
\def\He{\hbox{He}}

\newcount\eqnno
\def\sen{\the\secno}
\def\eqn#1{\global\advance\eqnno by 1
      \eqno(\the\eqnno)
      \expandafter \xdef\csname #1\endcsname {\the\eqnno}\relax }
\def\eqnn#1{\global\advance\eqnno by 1
      (\the\eqnno)
      \expandafter \xdef\csname #1\endcsname {\the\eqnno}\relax }
\def\eqnm#1#2{\global\advance\eqnno by 1
      (\the\eqnno\hbox{#2})
      \expandafter \xdef\csname #1\endcsname {\the\eqnno}\relax }
\def\eqnr#1{(\the\eqnno\hbox{#1})}

\newcount\caseno
\def\case#1{\vskip-\lastskip
  \vskip4pt plus2pt minus2pt
  \bigbreak \global\advance\caseno by 1
\noindent\underbar{{\sc Case}
\the\secno.\the\subno.\the\caseno}\enskip{#1}. }

\def\Case#1{\vskip-\lastskip
  \vskip4pt plus2pt minus2pt
  \bigbreak \global\advance\caseno by 1
\noindent {\sc
Case}\ \the\secno.\the\subno.\the\exno{\romannumeral\the\caseno}
\enskip \underbar{#1}.\quad}

\def\p{Pain\-lev\'e}
\def\Pi{first \p\ equation}

\font\sevensy=cmsy7

\font\sevenrm=cmr7
\font\sevenbf=cmbx7

\def\cite#1{[#1]}

\def\widehat{\mathaccent"362}
\def\widetilde{\mathaccent"365}
\def\~#1{{\bf\widetilde{\mit#1}}}
\def\^#1{{\bf\widehat{\mit#1}}}

\def\hide#1{}

\newcount\refno
\refno=0

\def\refdpp#1#2#3#4{\global\advance\refno by 1
\expandafter \xdef\csname #1\endcsname {\the\refno}\relax}{}{}{}
\def\refjl#1#2#3#4#5#6{\global\advance\refno by 1
\expandafter \xdef\csname #1\endcsname {\the\refno}\relax}{}{}{}{}{}{}
\def\refbk#1#2#3#4#5{\global\advance\refno by 1
\expandafter \xdef\csname #1\endcsname {\the\refno}\relax}{}{}{}{}
\def\refcf#1#2#3#4#5#6#7{\global\advance\refno by 1
\expandafter \xdef\csname #1\endcsname {\the\refno}\relax}{}{}{}{}{}{}

\refjl{refRGH}{A. Ramani, B. Grammaticos and J. Hietarinta}
{Phys. Rev. Lett.}{67}{1829}{1991}
\refjl{refRGV}{A. Ramani,\ B. Grammaticos and V. Papageorgiou}
{Phys. Rev. Lett.}{67}{1825}{1991}
\refbk{refInce}{E.L. Ince}{Ordinary Differential Equations}{Dover, New
York}{1956}
\refjl{refASa}{M.J. Ablowitz and H. Segur}{Phys. Rev.
Lett.}{38}{1103}{1977}
\refbk{refAC}{M.J. Ablowitz and P.A. Clarkson}{Solitons, Nonlinear
Evolution Equations and Inverse Scattering, {L.M.S. Lect. Notes Math.},
Vol. {149}}{CUP, Cambridge}{1991}
\refjl{refGromak}{V.I. Gromak}{Diff. Eqns.}{14}{1510}{1977}
\refjl{refBK}{E. Brezin and V. Kazakov}{Phys. Lett. B}{236}{144}{1990}
\refjl{refGM}{D.J. Gross and A.A. Migdal}{Phys. Rev.
Lett.}{64}{127}{1990}
\refjl{refDoug}{M.R. Douglas}{Phys. Lett. B}{238}{176}{1990}
\refjl{refFIK}{A.S. Fokas, A.R. Its and A.V. Kitaev}{Commun. Math.
Phys.}{142}{313}{1991}
\refjl{refPS}{V. Periwal and D. Shewitz}{Phys. Rev.
Lett.}{64}{1326}{1990}
\refjl{refNP}{F.W. Nijhoff and V. Papageorgiou}{Phys. Lett.
A}{153}{337}{1991}
\refjl{refFGR}{A.S. Fokas, B. Grammaticos and A. Ramani}{J. Math.
Anal.
Appl.}{180}{342}{1993}
\refcf{refGR}{B. Grammaticos and A. Ramani}{Kluwer, Dordrecht,
1993}{Applications of Analytic and Geometric Methods to Nonlinear
Differential
Equations}{ed. P.A. Clarkson}{{NATO ASI Series C: Mathematical and
Physical Sciences\/}, Vol. {413}}{299}
\refjl{refKOSGR}{K. Kajiwara, Y. Ohta, J. Satsuma, B. Grammaticos and
A.
Ramani}{J. Phys. A: Math. Gen.}{27}{915}{1994}
\refjl{refGNPRS}{B. Grammaticos, F.W. Nijhoff, V. Papageorgiou, A.
Ramani and J. Satsuma}{Phys. Lett. A}{185}{446}{1994}
\refjl{refPNGR}{V. Papageorgiou, F.W. Nijhoff, B. Grammaticos and A.
Ramani}{Phys. Lett. A}{164}{1825}{1992}
\refdpp{refBCHI}{A.P. Bassom,\ P.A. Clarkson\ and A.C.
Hicks}{``B\"acklund
transformations and solution hierarchies for the fourth \p\ equation'',
Stud.
Appl. Math., to appear}{1994}
\refdpp{refBCHII}{A.P. Bassom,\ P.A. Clarkson\ and A.C. Hicks}
{``On the application of solutions of the fourth Painlev\'e equation to
various
physically motivated nonlinear partial differential
equations'', preprint {\bf M94/32}, Department of Mathematics,
University of
Exeter}{1994}
\refjl{refMurata}{Y. Murata}{Funkcial. Ekvac.}{28}{1}{1985}
\refjl{refTGR}{K.M. Tamizhmani,\ B. Grammaticos and A. Ramani}
{Lett. Math. Phys.}{29}{49}{1993}
\refbk{refASte}{M. Abramowitz and I.A. Stegun}{Handbook of Mathematical
Functions}{Dover, New York}{1965}
\refjl{refLukash}{N.A. Lukashevich}{Diff. Eqns.}{3}{395}{1967}
 \refjl{refBCHM}{A.P. Bassom, P.A. Clarkson, A.C. Hicks and J.B.
 McLeod}{Proc. R.
Soc. Lond. A}{437}{1}{1992}
 \refjl{refBCH}{A.P. Bassom, P.A. Clarkson and A.C. Hicks}{IMA J.
 Appl.
Math.}{50}{167}{1993}

\title{New exact solutions of the discrete}{fourth Painlev\'e equation}
\author{Andrew P.\ Bassom\footnote{$^1$}{\rm Email:
drew@maths.exeter.ac.uk}
and Peter A.\ Clarkson\footnote{$^2$}{\rm Email:
clarkson@maths.exeter.ac.uk}}
\address{Department of Mathematics, University of Exeter, Exeter, EX4
4QE, U.K.}

\beginabstract
\bigskip\medskip
In this paper we derive a number of exact solutions of the discrete
equation
$$x_{n+1}x_{n-1}+x_n(x_{n+1}+x_{n-1})=
{-2z_nx_n^3+(\eta-3\delta^{-2}-z_n^2)x_n^2+\mu^2\over
(x_n+z_n+\gamma)(x_n+z_n-\gamma)},\eqno(1)$$
where $z_n=n\delta$ and $\eta$, $\delta$, $\mu$ and $\gamma$ are
constants.
In an appropriate limit (1) reduces to the
fourth \p\ (PIV) equation
$${\d^2w\over\d z^2} = {1\over2w}\left({\d w\over\d
z}\right)^2+\tfr32w^3 +
4zw^2 + 2(z^2-\alpha)w +{\beta\over w},\eqno(2)$$
where $\alpha$ and $\beta$ are constants and (1) is commonly referred
to as
the discretised fourth Painlev\'e equation. A suitable factorisation of
(1) facilitates the identification of a number of solutions which take
the form of ratios of two polynomials in the variable $z_n$. Limits of
these solutions yield rational solutions of PIV (2). It is also known
that there
exist exact solutions of PIV (2) that are expressible in terms of the
complementary error function and in this article we show that a
discrete
analogue of this function can be obtained by analysis of (1).

\endabstract
\vfill
\date
\eject

\headline={\ifodd\pageno\rightheadline\else\leftheadline\fi}
\def\rightheadline{\tenrm\hfil {\eightrm Andrew P.\ Bassom
and Peter A.\ Clarkson
}\hfil\folio}
\def\leftheadline{\tenrm\hfil {\eightrm Exact solutions of
the discrete fourth \p\ equation}\hfil\folio}

\baselineskip=13.5pt

\section{Introduction}
The discrete fourth \p\ equation (d-PIV) was recently derived by
Ramani,
Grammaticos \& Hietarinta [\refRGH] using the method of singularity
confinement
[\refRGV]. It is given by the three-point, non-autonomous mapping
$$x_{n+1}x_{n-1}+x_n(x_{n+1}+x_{n-1})=
{-2z_nx_n^3+(\eta-3\delta^{-2}-z_n^2)x_n^2+\mu^2\over
(x_n+z_n+\gamma)(x_n+z_n-\gamma)},\eqn{eqI}$$
with the variable $x_n$ to be found in terms of $z_n\equiv
n\delta+\zeta$.
This mapping is identified as the discretised version of the continuous
fourth
\p\ equation (PIV)
$${\d^2w\over\d z^2} = {1\over2w}\left({\d w\over\d
z}\right)^2+\tfr32w^3 +
4zw^2 + 2(z^2-\alpha)w +{\beta\over w},\eqn{eqII}$$ where $\alpha$ and
$\beta$
are arbitrary constants. This is observed by taking the limit of (\eqI)
as
$\delta\to 0$, with $\gamma=1/\delta$ and $\eta$ and $\mu$ finite. This
process yields (\eqII) with the parameters $\alpha$ and $\beta$ in that
equation
related to $\eta$ and $\mu$ according to
$$\alpha=\tfr14\eta\qquad\qquad\hbox{and}\qquad\quad\beta=-\tfr12\mu^2.
\eqno\eqnm{eqIII}{a,b}$$

The six (continuous) \p\ equations (PI--PVI) were first derived around
the turn
of the century in an investigation by \p\ and his colleagues into which
second-order ordinary differential equations have the property that the
singularities other than poles of any of the solutions are independent
of the
particular solution and so are dependent only upon the equation (cf.,
[\refInce]); this property is now known as the \p\ property. There has
been
considerable interest in \p\ equations over the last few years
primarily due to
the fact that they arise as reductions of soliton equations solvable by
inverse
scattering as first demonstrated by Ablowitz and Segur [\refASa].
Although
first discovered from strictly mathematical considerations, the
\p\ equations
have appeared  in various of physical applications (cf., [\refAC] and
the
references therein). The \p\ equations may also be thought of as
nonlinear analogues of the classical special functions though they are
known to
be transcendental since their solution is not expressible in terms of
elementary
functions. However rational solutions and one-parameter families of
solutions
of the \p\ equations are expressible in terms of special functions are
known to
exist for particular values of the parameters. For example, there exist
solutions of PII, PIII and PIV that are expressed in terms of Airy,
Bessel and
parabolic cylinder functions, respectively (cf., [\refGromak]).

Recently there has been considerable interest in integrable mappings
and
discrete systems, including discrete analogues of the \p\ equations.
Some of
these mapping and discrete equations arise in physical applications.
For
example, a discrete analogue of PI (d-PI) arose in the study of the
partition
function in a two-dimensional model of quantum gravity
[\refBK--\refFIK].
Subsequently a discrete analogue of PII (d-PII) was derived in
[\refPS,\refNP]
and later discrete analogues of PIII--PV (d-PIII--d-PV) were obtained
[\refRGH]; for further details on the derivation of the discrete
\p\ equations
see, for example, [\refFGR,\refGR]. One important result of the
investigations
is that the form of the discrete \p\ equations is not unique since
there exist
several possible discrete analogues of the \p\ equations. Kajiwara
\etal
[\refKOSGR] and Grammaticos \etal [\refGNPRS] have derived exact
solutions of
d-PII and d-PIII in terms of discrete Airy and discrete Bessel
functions,
respectively, in analogue to the aforementioned results for the
associated
continuous \p\ equations. We further remark that Lax pairs and
isomonodromic
deformation problems are known to exist for d-PI [\refFIK], d-PII
[\refNP] and
d-PIII [\refPNGR]. However, at present, there is no discrete analogue
of PVI,
nor are there Lax pairs for the versions of d-PIV and d-PV derived in
[\refGNPRS].

In recent work [\refBCHI,\refBCHII], we have been concerned with the
investigation of \bts\ and exact solutions for PIV (\eqII) together
with an examination of various applications of these solutions to
several
physically motivated nonlinear partial differential equations. The
purpose of
the present article is to discuss some new solutions of d-PIV (\eqI).
In
[\refBCHI] we demonstrated how all known exact solutions of (\eqII) can
be
categorised into one of three families; in two of these solutions can
be
determined in terms of the complementary error and parabolic cylinder
functions
whilst the third family consists of solutions which can be expressed as
the
ratio of two polynomials in $z$.  Since d-PIV reduces to (\eqII) in the
appropriate limit then it can be expected that exact solutions of
(\eqI) exist
which should tend to known continuous solutions in the same limit. In
\S2 we
shall principally concern ourselves with rational solutions of d-PIV
and in \S3
we identify solutions of (\eqI) which can be thought of as
discrete analogues of the complementary error function hierarchy
discussed in
[\refBCHI]. In \S4 we make a few closing remarks.

\section{Rational solutions of d-PIV}
Before we start our study of rational solutions of (\eqI) it is
convenient
to summarise the state of knowledge of rational solutions for
PIV (\eqII). It is easily verified that $w=1/z$ satisfies (\eqII)
with the parameter choices $\alpha=2$, $\beta=-2$ and this is a simple
example of a rational solution of this equation. Two families of such
solutions take the forms
$$w(z;\alpha,\beta)={P_{n-1}(z)\over Q_n(z)},\qquad\quad\qquad\quad
 w(z;\alpha,\beta)=-2z+{P_{n-1}(z)\over Q_n(z)},\eqno\eqnm{eqIV}{a,b}$$
where $P_m(z)$ and $Q_m(z)$ denote some polynomials of degree $m$
consisting
of either entirely even or else entirely odd powers of $z$.
Murata [\refMurata] has shown
that (\eqII) admits unique rational solutions of type (\eqIV{a}) or
(\eqIV{b})
if the parameters are of the form

$$(\alpha,\beta)=\left( \pm k,-2(1+2n+k)^2\right);\qquad k, n\in\Z,
\quad n\leq -1,\quad k\geq -2n,\eqno\eqnm{eqV}{a}$$
and
$$(\alpha,\beta)=\left( k, -2(1+2n+k)^2\right);\qquad k, n\in\Z,\quad
n\geq 0,\quad k\geq -n,\eqno\eqnr{b}$$
respectively. A further family of rational solutions of PIV is
characterised by
$$w(z;\alpha,\beta)=-\tfr23z+{P_{n-1}(z)\over Q_n(z)},\eqn{eqVI}$$
where $(\alpha,\beta)=\left( n_1,-\tfr29(1+3n_2)^2\right)$ with
$n_1$ and $n_2$ either both even or both odd integers. An extended
discussion of the important properties of these rational solution
families
together with tables containing the first few solutions in each of
these
three hierarchies are to be found in [\refBCHI].

Tamizhmani, Grammaticos \& Ramani [\refTGR] noted that if we set
$$\eta=3\delta^{-2}+\gamma^2-2(a^2+b^2),\qquad\quad\mu=a^2-b^2,$$
then the d-PIV equation (\eqI) can be factorised as
$$(x_{n+1}+x_n)(x_{n-1}+x_n)={(x_n+a+b)(x_n+a-b)(x_n-a+b)(x_n-a-b)
\over (x_n+z_n+\gamma)(x_n+z_n-\gamma)},\eqn{eqVII}$$
and the parameters $\alpha$ and $\beta$, as defined by (\eqIII), are
now
given by
$$\alpha=\tfr34\delta^{-2}+\tfr14\gamma^2-\tfr12(a^2+b^2),\qquad
\beta=-\tfr12(a^2-b^2)^2.\eqn{eqVIII}$$
Simple rational solutions of (\eqVII) can be found with $x_n$
proportional to $z_n$; these are
$$ x_n=-2z_n,\qquad a=\tfr12\delta+\gamma,\quad
 b=\tfr12\delta-\gamma,\quad \alpha=\tfr34
\left(\delta^{-2}-\gamma^2\right)-
\tfr14\delta^2,\quad\beta=-2\gamma^2\delta^2,\eqn{eqIXa}$$
and
$$ x_n=-\tfr23z_n,\qquad a=\tfr16\delta+\gamma,\quad
 b=\tfr16\delta-\gamma,\quad \alpha=\tfr34\left(\delta^{-2}-\gamma^2
\right)-\tfr1{36}\delta^2,\quad
\beta=-\tfr29\gamma^2\delta^2.\eqn{eqIXb}$$
In the limit as $\delta\to 0$, with $\gamma=1/\delta$, these discrete
solutions tend to
$$w(z;0,-2)=-2z,\qquad w(z;0,-\tfr29)=-\tfr23z,$$
respectively, which are the first members
 of the PIV hierarchies typified by (\eqIV{b}) and
(\eqVI). If solutions of (\eqVII) proportional to $1/z_n$ are sought
then
it is found that
$$x_n=-{\delta(\delta\pm\gamma)\over z_n},\qquad
a=\tfr32\delta\pm\gamma,
\quad b=\tfr12\delta\pm\gamma,\quad\alpha=\delta^{-2}-\gamma^2
\mp 2\gamma\delta-\tfr54\delta^2,
\quad \beta=-2\delta^2(\delta\pm\gamma)^2,\eqn{eqX}$$
and so,  in the limit as  $\delta\to 0$, with $\gamma\delta=1$,
we have $x_n\to w(z;\pm 2,-2)=\pm 1/z$ and these
continuous solutions are members of the family described by
(\eqIV{a}).

More complicated rational solutions of d-PIV (\eqI) can
be deduced by rewriting (\eqVII)
as the pair
$$\eqalignno{&x_n+x_{n-1}={ (x_n+a+b)(x_n+a-b)\over (x_n+z_n\pm
\gamma)},
&\eqnm{eqXI}{a}\cr
&x_n+x_{n+1}={ (x_n-a+b)(x_n-a-b)\over (x_n+z_n\mp \gamma)}.
&\eqnr{b}\cr}$$
Tamizhmani \etal\ [\refTGR] speculated that such a formalism ought to
lead
to discrete rational solutions though they did not present any such
solutions.
It is a routine calculation to verify that equations (\eqXI{a}) and
(\eqXI{b}) are compatible if and only if
$$a=\tfr12\delta\pm\gamma.\eqn{eqXII}$$
[We remark at this stage that (\eqXI) is not the only possibility for
the
splitting of (\eqVII). Easy generalisations of (\eqXI) include the
multiplication of the right hand sides of (\eqXI{a}) and (\eqXI{b}) by
constants $C$ and $1/C$ respectively or the taking of the factors in
different
pairings. However, it can be shown that in either of these cases we
obtain an incompatible couple of equations so that the choice of
factorisation (\eqXI) is not as specialised and restrictive as it might
first
appear.]

If the condition (\eqXII) is satisfied then we can seek solutions of
(\eqI)
by finding solutions of the simpler form (\eqXI{b}) (or (\eqXI{a})
would
do equally well). If we write $x_n=P_n/Q_n$ then (\eqXI{b}) can be
recast
as
$$\eqalignno{&\kappa_{n+1}P_{n+1} =(z_n\pm\gamma+\delta)P_n-\mu Q_n,
&\eqnm{eqXIII}{a}\cr
       &\kappa_{n+1}Q_{n+1} =-(z_n\mp\gamma)Q_n-P_n, &\eqnr{b}\cr}$$
where the `separation' parameter $\kappa_{n+1}$ can depend both on $n$
and $z_n$ (recall that $\mu=a^2-b^2$). However, if for the moment
we take $\kappa_{n+1}=1$, then eliminating $P_n$ between (\eqXIII{a})
and
(\eqXIII{b}), setting $\gamma=1/\delta$ and taking the limit as
$\delta\to 0$
shows that $Q_n\to q(z)$ where $q(z)$ satisfies
where
$${{\d}^2q\over {\d}z^2}=(z^2+\mu-1)q.\eqno\eqnm{eqXIV}{a}$$
If in this equation we set $q(z)=\eta(\xi)$ with $\xi=\sqrt2\,z$ then
we obtain
$${{\d}^2\eta\over{\d}\xi^2}=(\tfr14\xi^2+\tfr12\mu-\tfr12)
\eta,\eqno\eqnr{b}$$
which is the parabolic cylinder equation (cf., [\refASte]).
It is well known that when $\mu=-2n$ this equation admits the solution
$$\eta(\xi)=\He_n(\xi)\exp\left(-\tfr14\xi^2\right),\eqno\eqnm{eqXV}{a}$$
where $\He_n(\xi)$ is the Hermite polynomial of degree $n$ given by
$$\He_n(\xi)=(-1)^n\exp\left(\tfr12\xi^2\right){{\d}^n\over {\d}
\xi^n}\left[\exp\left(-\tfr12\xi^2\right)\right].\eqno\eqnr{b}$$

If the variable $Q_n$ is eliminated from equations (\eqXIII) and the
usual limit taken then it follows in a manner similar to that outlined
above
that rational solutions $x_n=P_n/Q_n$ of (\eqI) exist which tend to a
function
of the form $-\He_m'(\xi)/\He_m(\xi)$ whenever $\mu=-2m$, with
$m\in\Na$.
It has been long established that solutions of (continuous) PIV
(\eqII) can be related to Hermite functions whenever the
parameters $\alpha$ and $\beta$ take certain values. In particular,
Lukashevich [\refLukash] proved that
$$\eqalignno{ w(z;-(\nu+1),-2\nu^2)&=-{1\over\phi_{\nu}}{{\d}\phi_{\nu}
\over{\d}z},&\eqnm{eqXVI}{a}\cr
 w(z;-\nu,-2(\nu+1)^2)&=-2z-{1\over\phi_{\nu}}{{\d}\phi_{\nu}
\over{\d}z},&\eqnr{b}\cr
w(z;\nu,-2(\nu+1)^2)&=-2z-{1\over\psi_{\nu}}{{\d}\psi_{\nu}
\over{\d}z},&\eqnr{c}\cr
w(z;\nu+1,-2\nu^2)&={1\over\psi_{\nu}}{{\d}\psi_{\nu}
\over{\d}z},&\eqnr{d}\cr}$$
where $\phi_{\nu}(z)$ and $\psi_{\nu}(z)$ are any solutions of the
Weber-Hermite equations
$$\eqalignno{ {{\d}^2\phi_{\nu}\over{\d}z^2}-2z{{\d}\phi_{\nu}
\over{\d}z} +2\nu\phi_{\nu}&=0,&\eqnm{eqXVII}{a}\cr
{{\d}^2\psi_{\nu}\over{\d}z^2}+2z{{\d}\psi_{\nu}
\over{\d}z} -2\nu\psi_{\nu}&=0.&\eqnr{b}\cr}$$
If $\nu=n$, with $n$ a positive integer, then $\phi_n(z)$ and
$\psi_n(z)$ are
polynomials of degree $n$ given by
$$\phi_n(z)=(-1)^n\exp\left(z^2\right){{\d}^n\over {\d}
z^n}\left[\exp\left(-z^2\right)\right],\qquad
\psi_n(z)=\exp\left(-z^2\right){{\d}^n\over {\d}
z^n}\left[\exp\left(z^2\right)\right].$$
(These are also expressible in terms of the Hermite polynomial
$\He_n(\xi)$
defined above.)

Motivated by these comments concerning the continuous PIV case,
we return to equations (\eqXIII) for the discrete situation. If we
write
$$(P_n,Q_n)=(A_n,B_n)\times (-\delta)^n\Gamma\left( {z_n-m\delta
\mp\gamma\over\delta}\right),\eqn{eqXVIII}$$
where $\Gamma(z)$ denotes the usual Gamma function, and choose
$\kappa_{n+1}\equiv 1$, then we obtain the pair
$$\eqalignno{
   &(z_n-m\delta\mp\gamma)A_{n+1}+(z_n+\delta\pm\gamma)A_n=\mu B_n,
&\eqnm{eqXIX}{a}\cr
   &(z_n-m\delta\mp\gamma)B_{n+1}-(z_n\mp\gamma)B_n=A_n.&\eqnr{b}\cr}$$
For $m\in\N$ we can find exact solutions of these equations with
$A_n$ and $B_n$ taking the forms of polynomials in $z_n$, consisting
of either only even or only odd powers, and of degrees $m-1$ and $m$
respectively. The first few solutions in this hierarchy are
\def\de{\delta}
\def\ga{\gamma}
$$\eqalignno{
   m=1,\qquad&\qquad x_n=-{\delta(\delta\pm\gamma)\over
   z_n},&\eqnm{eqXX}{a}\cr
   m=2,\qquad&\qquad
   x_n=-{2\de(3\de\pm2\ga)z_n\over2z_n^2-\de(2\de\pm\ga)},
&\eqnr{b}\cr
   m=3,\qquad&\qquad x_n=-{3\de(2\de\pm\ga)\left[2z_n^2-\de
(3\de\pm\ga)\right]\over z_n\left[ 2z_n^2-\de(8\de\pm 3\ga)\right] },
&\eqnr{c}\cr
   m=4,\qquad&\qquad x_n=-{4\de(5\de\pm 2\ga)\left[ 2z_n^2-\de
(11\de\pm 3\ga)\right]z_n\over 4z_n^4-4\de(10\de\pm 3\ga)z_n^2+3\de^2
(3\de\pm\ga)(4\de\pm\ga)},
&\eqnr{d}\cr
   m=5,\qquad&\qquad x_n=-{5\de(3\de\pm\ga)\left[ 4z_n^4-4\de(13\de
\pm 3\ga)z_n^2+3\de^2(3\de\pm\ga)(4\de\pm\ga)\right]\over
z_n\left[ 4z_n^4-20\de(4\de\pm\ga)+\de^2(256\de^2\pm 125\ga\de+15\ga^2)
\right] }.&\eqnr{e}
\cr}$$

In each case the corresponding value of $\mu$ in (\eqXIX{a}) is
$\mu=-\de m\left[ (m+1)\de \pm 2\ga\right]$ and this leads to the
respective
values of parameters $\alpha$ and $\beta$, as defined by (\eqVIII),
given by
$$\alpha=\tfr34(\de^{-2}-\ga^2)\mp
(m+1)\ga\de-\tfr14\de^2\left[2m(m+1)+1
\right],
\qquad \beta=-\tfr12m^2\delta^2\left[ (m+1)\de\pm 2\ga\right]^2.
\eqn{eqCVS}$$
We emphasise at this stage that the discrete solutions (\eqXX) are
exact and are valid for {\it any} $\de$ and $\ga$. We can recover
continuous
solutions by letting $\ga=1/\de$ and taking the limit as $\de\to 0$
which yields
$\alpha=\mp(m+1)$ and $\beta=-2m^2$. Then from (\eqXX) we obtain
$$\eqalignno{w(z;\mp 2,-2)&=\mp {1\over z},&\eqnm{eqXXI}{a}\cr
       w(z;\mp 3,-8)&=\mp {4z\over 2z^2\mp 1},&\eqnr{b}\cr
       w(z;\mp 4,-18)&= \mp {3(2z^2\mp 1)\over z(2z^2\mp 3)},
&\eqnr{c}\cr
       w(z;\mp 5,-32)&= \mp {8z(2z^2\mp 3)\over 4z^4\mp 12z^2
+3},&\eqnr{d}\cr
       w(z;\mp 6,-50)&=\mp {5(4z^4\mp 12z^2+3)\over z(4z^4\mp
20z^2+15)}.&\eqnr{e}\cr}$$
It is then clear that solutions (\eqXX) can be
thought of as discrete analogues of the (continuous) solutions of PIV
taking
forms given by (\eqXVI{a}) and (\eqXVI{d}), with $\nu=m$.

If in (\eqXIII) we let
$$(P_n,Q_n)=(A_n,B_n)\times \delta^n\thinspace\Gamma\left( {z_n-m\delta
\pm\gamma\over\delta}\right),\eqn{eqXXII}$$
then we obtain
$$\eqalignno{
   &(z_n-m\delta\pm\gamma)(A_{n+1}-A_n)=-\mu B_n,
&\eqnm{eqXXIII}{a}\cr
   &(z_n-m\delta\pm\gamma)B_{n+1}+(z_n\mp\gamma)B_n=-A_n.&\eqnr{b}\cr}$$
As previously, exact polynomial solutions of these can be derived for
integer
$m$ except that now $A_n$ and $B_n$ are of degrees $m+1$ and $m$
respectively.
Then the first few solutions of this type are
$$\eqalignno{ m=0,\qquad&\qquad x_n=-2z_n,&\eqnm{eqXXIV}{a}\cr
       m=1,\qquad&\qquad x_n=-{2z_n^2-\de(\de\mp\ga)\over z_n},
&\eqnr{b}\cr
       m=2,\qquad&\qquad x_n=-{2z_n\left[2z_n^2-\de(5\de\mp
3\ga)\right]\over 2z_n^2-\de(2\de\mp\ga)},
&\eqnr{c}\cr
       m=3,\qquad&\qquad x_n=-{4z_n^4-4\de(7\de\mp 3\ga)z_n^2
+3\de^2(3\de\mp\ga)(2\de\mp\ga)\over z_n\left[ 2z_n^2-\de(8\de\mp
3\ga)\right] },&\eqnr{d}\cr
       m=4,\qquad&\qquad x_n=-{2z_n\left[ 4z_n^4-20\de
(3\de\mp\ga)z_n^2+\de^2(146\de^2\mp 95\ga\de+15\ga^2)\right] \over
4z_n^4-4\de(10\de\mp 3\ga)z_n^2+3\de^2(4\de\mp\ga)(3\de\mp\ga)}.
&\eqnr{e}\cr}$$
The corresponding values of the parameters $\alpha$ and $\beta$ are
given
by
$$\alpha=\tfr34(\de^{-2}-\ga^2)\pm m\ga\de-\tfr14\de^2\left[2m(m+1)+
1\right],\qquad
 \beta=-\tfr12(m+1)^2\de^2(m\de\mp 2\ga)^2\eqn{eqasasd}$$
and in the limit $\de\to 0$ with $\ga=1/\de$,
solutions (\eqXXIV) reduce to the
functions
$$\eqalignno{ w(z;0,-2) &=-2z,&\eqnm{eqXXV}{a}\cr
	w(z;\pm 1,-8)&=-2z\mp{1\over z},&\eqnr{b}\cr
	w(z;\pm 2,-18)&=-2z\mp{4z\over 2z^2\pm 1},&\eqnr{c}\cr
	w(z;\pm 3,-32)&=-2z\mp{3(2z^2\pm 1)\over z(2z^2\pm 3)},
&\eqnr{d}\cr
	w(z;\pm 4,-50)&=-2z\mp{8z(2z^2\pm 3)\over 4z^4\pm 12z^2
+3}.&\eqnr{e}\cr}$$
These are all members of the so-called `$-2z$' hierarchy of rational
solutions for PIV (\eqII) and are of the form (\eqXVI{b}) or
(\eqXVI{c}), with $\nu=m$. It is straightforward to obtain further
exact rational solutions of d-PIV by
solving the pairs (\eqXIX) or (\eqXXIII) for higher values of $m$.

We note here that although we have found some rational solutions of
d-PIV (\eqI) it can be anticipated that further such solutions exist
which are
not derivable using the procedure described above. This deduction
follows from the observation that rational solutions of PIV (\eqII)
 are possible for all parameter values as described by
(\eqV) and that so far we have discrete solutions which, in the
appropriate limit, tend to the continuous solutions of type (\eqXVI).
Therefore, there should be discrete counterparts to the remaining
continuous rational solutions. These can be constructed by appealing to
some \bts\ for d-PIV which were given by Tamizhmani \etal\
[\refTGR]. These authors presented a sequence of transformations
which, given a solution of (\eqVII) with parameters $a$, $b$ and
$\gamma$,
yields a further solution of the same equation but now with parameters
$a+\delta$, $b$ and $\gamma$.
In short, suppose that $x_n$ is a solution of d-PIV  with
parameters $a$, $b$ and $\gamma$. Then
$$\~y_n=-{x_nx_{n+1}+x_n(\~z_n+a)+x_n(\~z_n-a)+b^2-a^2\over
x_n+x_{n+1}},\eqn{eqback}$$
with $\~z_n= z_n+\tfr12\delta$ also satisfies d-PIV but now with
parameters $\~a=a+\tfr12\delta$, $\~b=\gamma$ and $\~\gamma=b$.
The subtlety with transformation (\eqback) is that $\~y_n$ is defined
on
a lattice of points which is offset by $\tfr12\delta$ from the
original. In
order to obtain a solution valid at points coinciding with the initial
lattice
it is therefore necessary to reapply (\eqback) which has the overall
effect of
raising the value of $a$  by $\delta$ whilst keeping
$b$ and $\gamma$
invariant. Clearly $M$ applications of this sequence
will increment $a$ to $a+M\delta$ and leaves the other two parameters
unchanged so that from
a starting solution with parameters
$$a=\tfr12\de\pm\ga,\qquad b=(N+\tfr12)\de\pm\ga,\quad
N=0,1,2,\dots\eqn{eqXXVI}$$
(which are the parameter values in (\eqI) appropriate to the family of
solutions whose first few members are listed in (\eqXX)) we can obtain
a solution with
$$a=(M+\tfr12)\de\pm\ga,\qquad b=(N+\tfr12)\de\pm\ga,
\eqn{eqXXVII}$$
or, in terms of the parameters $\alpha$ and $\beta$ as given by
(\eqVIII),
$$\eqalignno{
 \alpha&=\tfr34(\de^{-2}-\ga^2)\mp\ga\de(M+N+1)+\left[ \tfr12(M-N)
(M+N+1)-(M+\tfr12)^2\right]\delta^2,&\eqnm{eqXXVIII}{a}\cr
 \beta&=-\tfr12(M-N)^2\de^2\left[(M+N+1)\de\pm 2\ga\right]^2.
&\eqnr{b}\cr}$$

In the usual limit our discrete rational solutions will then tend to
continuous ones with associated parameters $\alpha=\mp(M+N+1)$,
$\beta=-2(M-N)^2$. Now $M$ and $N$ can be chosen so as to force these
parameters to coincide with the form (\eqV{a}) for any $k$ and $n$ in
the permissible ranges; in this way we have a mechanism for deducing
discrete analogues of all those rational solutions of PIV (\eqII)
which take the form $P_{n-1}(z)/Q_n(z)$.

A similar argument can be applied to the discrete solutions (\eqXXV)
and this demonstrates that given these results then suitable
application
of the \bts\ contained in [\refTGR] will generate another
set of exact solutions of (\eqI). These comprise the discrete
counterpart to the `$-2z$' rational hierarchy of
(\eqII) which itself is characterised by the parameter values given in
(\eqV{b}).

We remark here that of the three hierarchies of rational solutions for
PIV (\eqII), the simple procedures outlined above have yielded discrete
analogues of only two of these; no solutions corresponding to the
$-\tfr23z$
family (\eqVI) have been found. The reason for this is that although
the full
discrete equation (\eqVII) admits the solution $x_n=-\tfr23z_n$ with
$a=\tfr16\delta\pm\gamma$, the splitting of this equation into the two
Ricatti-like forms (\eqXI) gives a pair of equations whose
compatibility
requires that $a=\tfr12\de\pm\ga$. Thus, as noted by Tamizhmani \etal\
[\refTGR], this solution is not linearizable through the splitting
assumption
and thus it is unsurprising that the discrete analogue of the
$-\tfr23z$
hierarchy of solutions cannot be generated in this way. However, we
shall
now show how use of the \bt\ (\eqback) discussed above can be used to
increment
the values of $a$ or $b$ by
$\pm\delta$ and thus lead to some more exact solutions in the discrete
$-\tfr23z$ hierarchy. (It is noted that d-PIV (\eqVII) is invariant by
interchange of parameters $a$ and $b$ or by a sign change of either of
these.
Therefore, given that two applications of (\eqback) increases $a$ by
$\delta$
it is easy to see how a suitable changes in sign of $a$ and $b$ or the
swopping
of these parameters allows transformations to be found that increase or
decrease the values of $a$ or $b$ by integral multiples of
$\delta$.) A few examples of simpler solutions in this hierarchy are:
$$\eqalignno{
x_n&=-\tfr23z_n,&\eqnm{eqtry}{a}\cr
a&=\tfr16\delta\pm\gamma,\quad b=-\tfr16\delta\pm\gamma,\quad
\alpha=\tfr34\delta^{-2}-\tfr34\gamma^2-\tfr{1}{36}\delta^2,\quad
\beta=-\tfr29\gamma^2\delta^2,\cr\cr
x_n&=-\tfr23z_n-{\delta(\delta\mp3\gamma)\over 3z_n},&\eqnr{b}\cr
a&=\pm\gamma-\tfr56\delta,\quad b=\mp\gamma+\tfr16\delta,\quad
\alpha=\tfr34(\delta^{-2}-\gamma^2)\pm\gamma\delta-\tfr{13}{36}\delta^2,\quad
\beta=-\tfr29\delta^2(2\gamma\pm\delta)^2,\cr\cr
x_n&=-{2z_n(2z_n^2\mp 3\gamma\delta+\delta^2)\over 3(2z_n^2\pm 3\gamma
\delta-2\delta^2)},&\eqnr{c}\cr
a&=\pm\gamma-\tfr76\delta,\quad b=\mp\gamma+\tfr56\delta,\quad
\alpha=\tfr34(\delta^{-2}-\gamma^2)\pm2\gamma\delta-\tfr{37}{36}\delta^2,\quad
\beta=-\tfr29\delta^2(\gamma\mp\delta)^2,\cr\cr
  x_n&=-{2z_n(4z_n^4- 45\gamma^2\delta^2+\delta^4)\over
  3(2z_n^2-3\gamma
\delta-\delta^2)(2z_n^2+3\gamma\delta-\delta^2)},&\eqnr{d}\cr
  a&=\pm\gamma+\tfr56\delta,\quad  b=\mp\gamma+\tfr56\delta,\quad
\alpha=\tfr34(\delta^{-2}-\gamma^2)-\tfr{25}{36}\delta^2,\quad
\beta=-\tfr{50}9\gamma^2\delta^2, \cr\cr
x_n&=-{ 8z_n^6-4\delta(\delta\pm3\gamma)z_n^4-2\delta^2(38\delta^2\mp
69
\gamma\delta+27\gamma^2)z_n^2+9\delta^3(4\delta^3\mp
11\gamma\delta^2+10
\gamma^2\delta\mp 3\gamma^3)\over 3z_n\left[4z_n^2-4\delta(4\delta\mp
3\gamma)z_n^2\pm 9\gamma\delta^2(\delta\mp\gamma)\right] },&\eqnr{e}\cr
a&=\pm\gamma-\tfr76\delta,\quad b=\mp\gamma+\tfr{11}6\delta,\quad
\alpha=\tfr34(\delta^{-2}-\gamma^2)\pm
3\gamma\delta-\tfr{85}{36}\delta^2,\quad
\beta=\tfr29\delta^2(3\delta\mp 2\gamma)^2.{\hbox to 50pt{\hfill}}
\cr}$$
In the limit as $\delta\to 0$, with $\gamma\delta= 1$, the solutions
(\eqtry)
reduce to
$$\eqalignno{
 w(z;0,-\tfr29)&=-\tfr23z,&\eqnm{soltt}{a}\cr
 w(z;\pm1,-\tfr89)&=-\tfr23z\mp{1\over z},&\eqnr{b}\cr
 w(z;\pm2,-\tfr29)&=-\tfr23z\pm{4z\over 2z^2\pm 3},&\eqnr{c}\cr
 w(z;0,-\tfr{50}9)&=-\tfr23z\pm{24z\over (2z^2-3)(2z^2+3)},&\eqnr{d}\cr
 w(z;\pm3,-\tfr89)&=-\tfr23z\pm{3(4z^4\pm 4z^2+3)\over z(4z^4\pm
 12z^2-9)},
&\eqnr{e}\cr}$$
which belong to the $-\tfr23z$ hierarchy of rational solutions of PIV
(\eqII);
see Table 4.1.3 of [\refBCHI] for a more extensive list of such
solutions.
The method of generating the $-\tfr23z_n$ discrete rational solutions
given
in (\eqtry) is intrinsically less satisfying than that employed for the
evaluation of solutions (\eqXX) and (\eqXXIV) in the other two
families;
this feature is a direct consequence of the fact that the
$-\tfr23z_n$ forms do not arise from a suitable factorisation of d-PIV
into a
pair of simple compatible equations akin to (\eqXI). Instead the
present method
is based on a direct implementation of the \bt\ (\eqback), which, for
the for
more  complicated solutions in the hierarchy, becomes an increasingly
laborious
task. However, we believe that the \bt\ (\eqback) will inevitably be
needed to
make a complete evaluation of each of the hierarchies and provide a
systematic
and efficient procedure for doing this.

\section{Discrete Complementary Error Function Solutions}
In \S2 we have deduced several new rational solutions
of d-PIV (\eqI). It has been known for a considerable time that PIV
(\eqII)
admits a number of other exact solutions which can be categorised into
separate
families. In addition to the rational hierarchy, solutions can be found
which are expressible in terms of the complementary error function and,
for certain half-integer and integer values of the parameters $\alpha$
and
$\beta$, in terms of parabolic cylinder functions. We note the close
connection
which exists between the complementary error function and rational
solution
hierarchies; solutions in the former can be found in terms of $\Psi(z)$
where
$$\Psi(z)={\psi'(z)\over\psi(z)},\qquad\qquad \psi (z)=1- C\,\erfc(z)
\equiv1-{2C\over\sqrt\pi}\int_z^{\infty}\exp(-t^2)\,{\d}t,
\eqno\eqnm{eqXXIX}{a,b}$$%
\hide{where
$$\erfc(z)={2\over\sqrt\pi}\int_z^{\infty}\exp(-t^2)\,{\d}t,\eqno\eqnr{c}$$}
with $C$ an arbitrary constant. When we choose $C=0$ then $\Psi\equiv
0$ and the complementary error solutions reduce to the rational forms
whose discrete analogues have been described above. The formulation
given
in \S2 enables us to deduce a discretised version of $\Psi(z)$ and in
order
to do this we return to system (\eqXIII). If we take $\mu=0$ there and
select $\kappa_{n+1}=-(z_n+\gamma)$ then
$$\eqalignno{
    & (z_n+\gamma)P_{n+1} +(z_n-\gamma+\delta)P_n=0,&\eqnm{eqXXX}{a}\cr
    & (z_n+\gamma)\left[ Q_{n+1}-Q_n\right]-P_n=0.&\eqnr{b}\cr}$$
If we choose $z_n=n\de$ then
$$P_{n+1}=\left[ {\gamma-(n+1)\de\over\gamma+n\de}\right]P_n\qquad
\hbox{and}\qquad Q_{n+1}-Q_n={P_n\over
\gamma+n\de}.\eqno\eqnm{eqXXXI}{a,b}$$
The first of these is easily solved to yield
$$P_n=\displaystyle{ \left[ \Gamma\left({\ga/\de}\right)\right]^2\over
\Gamma\left(-n+{\gamma/\delta}\right)
\Gamma\left(n+{\gamma/\delta}\right)}\, P_0\eqn{eqXXXII}$$
and it is clear that this is an even function in $n$. The continuous
limit
of (\eqXXXII) is best obtained by setting $\gamma=1/\de$, $\de=z/n$
and noting that
$$\ln (P_n/P_0)=\sum_{r=1}^{n}\thinspace
\ln\left(1-{rz^2\over
n^2}\right)-\sum_{r=1}^{n-1}\thinspace\ln\left(1+{rz^2\over
n^2}\right).\eqn{eqXXXIII}$$
Hence it follows that
$$\lim_{n\to\infty}\left\{\ln (P_n/P_0)\right\}=-z^2+O\left(
{1/n}\right),$$
and so  $$P_n\to P_0\exp(-z^2)\equiv P_*(z),$$ say. Consequently
(\eqXXXII)
can be thought of as a discrete version of $\psi'(z)\propto\exp(-z^2)$
and so the solution
of (\eqXXX{b}) for $Q$ should yield an appropriate discretised form for
$\psi(z)$ which by (\eqXXIX{b}) is closely related to the
error function $\erfc(z)$. An explicit solution of equation (\eqXXX{b})
is not immediately apparent, though it can readily be seen that
$$Q_{n+1}=\sum_{j=0}^n{P_j\over \gamma+j\de }\thinspace +\thinspace
Q_0,
\eqn{eqXXXIV}$$
where $Q_0$ is a constant and the
$P_j$ are as in (\eqXXXII).
This sum is not straightforward to evaluate but for the
present purposes it is adequate to merely note that as $\de\to 0$, with
$\ga=1/\de$, then
$$Q_{n+1}\longrightarrow\thinspace\thinspace
\de\sum_{j=0}^n{P_j}\thinspace +\thinspace Q_0,
$$
which in turn asymptotes to the integral of $P_*(z)$; i.e.\ to a
multiple of the complementary error function plus a constant. Thus in
forming the solution $x_n=P_n/Q_n$ we have a discrete equivalent to the
continuous PIV solution $w(z;1,0)=\Psi(z)$ defined above. Solutions in
the (continuous) complementary error function hierarchy are of the form
$$w(z;\alpha,\beta)={P(z,\Psi(z))\over Q(z,\Psi(z))},$$
where $P$ and $Q$ are polynomials in $z$ and $\Psi(z)$, and exist for
precisely
those parameter values given in (\eqV). What we have developed here is
the
discrete analogue of the simplest solution in the complementary error
function
hierarchy; furthermore, this is the discrete analogue of the most
elementary of
the ``bound-state'' solutions of PIV derived in [\refBCHM] (see also
[\refBCH]),
which themselves form a special case complementary error function
hierarchy.

This technique can be adapted to find discrete counterparts to further
exact solutions within the complementary error function hierarchy.
If we take
$$\eqalignno{\alpha^{(m)}&=\tfr34\left(\delta^{-2}-\gamma^2\right)+
(m+1)\ga\de-\tfr14\left[2m(m+1)+1\right]\de^2,&\eqnm{eqXXXVIII}{a}\cr
\beta^{(m)}&=-\tfr12m^2\de^2\left[ 2\ga
-(m+1)\de\right]^2,&\eqnr{b}\cr}$$
for $m=0,1,\dots$, and also write the corresponding discretised
solution $x_n^{(m)}=P_n^{(m)}\!/Q_n^{(m)}$, then equations (\eqXIII),
with the
lower sign taken, may be written as
$$\eqalignno{&\kappa_{n+1}^{(m)}P_{n+1}^{(m)} =
(z_n-\gamma+\delta)P_n^{(m)}-\mu^{(m)} Q_n^{(m)},
&\eqnm{eqXXXIX}{a}\cr
&\kappa_{n+1}^{(m)}Q_{n+1}^{(m)} =-(z_n+\gamma)Q_n^{(m)}-P_n^{(m)},
&\eqnr{b}\cr}$$ where $\mu^{(m)}=m\de\left[ 2\ga-(m+1)\de\right]$. In
the limit
as $\de\to 0$, with
$\ga\de=1$, it is clear from (\eqXXXVIII) that we will obtain solutions
with $\alpha=m+1$, $\beta=-2m^2$ and it is the discrete complementary
error function solution with index $m=0$ that we have derived in
(\eqXXX)-(\eqXXXIV) above. If we choose the separation variable
$\kappa_{n+1}^{(m)}=-(z_n+\ga-m\de)$ then by cross-elimination we can
show from (\eqXXXIX) that the variables $P_n^{(m)}$ and $Q_n^{(m)}$
satisfy
$$\eqalignno{&[z_n+(\ga-\de)-m\de]P_{n+2}^{(m)}-2(\ga-\de)P_{n+1}^{(m)}
-[z_n-(\ga-\de)+m\de]P_n^{(m)}=0,&\eqnm{eqXL}{a}\cr
&[z_n+\ga-(m+1)\de]Q_{n+2}^{(m)}-2\ga Q_{n+1}^{(m)}-
[z_n-\ga+(m+1)\de]Q_n^{(m)}=0.&\eqnr{b}\cr}$$
Now it can be seen that if for some particular choice $m=M$ we write
$\^\ga=\ga-\de$ in (\eqXL{a}) then the governing equation for
$P_n^{(M)}$
is identical to that for $Q_n^{(M-1)}$ with $\ga$ in (\eqXL{b}) written
as
$\^\ga$. The result of this observation is that given the solution
$Q_n^{(M-1)}$, we can immediately deduce the form of $P_n^{(M)}$ simply
by
replacing all occurrences of $\ga$ in the former solution with
$\ga-\de$.
Furthermore, since the continuous limit of our discretised solutions is
obtained
by letting $\de\to 0$ and $\ga\de\to 1$, it is clear that the
appropriate limit
for $P_n^{(M)}$ is precisely the same as that for $Q_n^{(M-1)}$. Thus,
in
practice, if we have solutions $P_n^{(M-1)}$ and $Q_n^{(M-1)}$ then
$P_n^{(M)}$
is directly derived from $Q_n^{(M-1)}$ and $Q_n^{(M)}$ can be obtained
by
solving (\eqXXXIX{b}).

The discretised complementary function solutions described here are
exact
in the sense that they are precise solutions of the d-PIV equation
(\eqVII) but the practical difficulty with them is that the derivation
of
explicit formulae for the solutions is distinctly awkward. The main
reason
for this stems from the fact that a closed form for the $Q_n^{(0)}$
solution
given by (\eqXXXIV) is  unknown and then, because of the nature of
systematic manner for deriving further solutions that we have just
described,
this difficulty with the $Q_n^{(0)}$ form propagates itself through the
rest of the hierarchy. However, we shall demonstrate how one can obtain
the
next complementary error function solution and this shows how the
procedure
operates in practice. Using the notation suggested by (\eqXXXVIII) we
have
the results that
$$P_n^{(0)}=
\displaystyle{ \left[ \Gamma\left({\ga/\de}\right)\right]^2\over
\Gamma\left(-n+{\gamma/\delta}\right)
\Gamma\left(n+{\gamma/\delta}\right)} P_0^{(0)},\qquad
Q_{n+1}^{(0)}=Q_0^{(0)}+\sum_{j=0}^n {P_j^{(0)}\over \ga+j\de},
\eqno{\eqnm{eqXLI}{a,b}}$$ which we have noted tend to multiples of the
functions $\psi'(z)$ and $\psi(z)$ respectively (see (\eqXXIX)) in the
appropriate limit. We obtain $P_n^{(1)}$ immediately from (\eqXLI{b})
by
replacing $\ga$ by $\ga-\de$, since $P_n^{(1)}=Q_n^{(0)}$, and so
$$P_n^{(1)}=Q_0^{(0)}+\sum_{j=0}^{n-1}{P_j^{(0)}\over \ga+(j-1)\de}.
\eqno\eqnm{eqXLII}{a}$$
Then we obtain $Q_n^{(1)}$ by solving (\eqXXXIX{b}) which yields
$$Q_n^{(1)}=[\ga-(n-1)\de]\sum_{j=0}^{n-2}{P_j^{(1)}\over (\ga+j\de)[
\ga+(j+1)\de]} +{P_{n-1}^{(1)}\over
\ga+(n-2)\de}+Q_0^{(1)}.\eqno\eqnr{b}$$
We can readily see the difficulty in simplifying these solutions
further but
we can show from (\eqXLII{b}) that as $\de\to 0$ with $\ga\de=1$ so
$$Q_n^{(1)}\to\de\sum_{j=0}^{n-1}P_j^{(1)}+Q_0^{(1)}.\eqn{eqXLIII}$$
Since $P_n^{(1)}\to\psi(z)$ in this limit, so (\eqXLIII) indicates that
$Q_n^{(1)}$ asymptotes the integral of $\psi(z)$. Moreover, since
$\psi(z)$
satisfies
$${\d^2\psi\over\d z^2} +2z{\d\psi\over\d z}=0,$$ this then means that
$$Q_n^{(1)}\to z\psi(z)+\tfr12{\d\psi\over\d z}\eqn{eqXLIV}$$
and hence
$$x_n^{(1)}\equiv {P_n^{(1)}\over Q_n^{(1)}}\to {2\psi(z)\over
2z\psi(z)+\psi'(z)}= {2\over\Psi(z)+2z},\eqn{eqXLV}$$
where $\Psi(z)$ is as defined in (\eqXXIX{a}). Thus we conclude that
$x_n^{(0)}$
and $x_n^{(1)}$ provide discretised forms of the first two
complementary
error function solutions of the continuous PIV equation (\eqII)
with parameters as given
by (\eqXXXVIII), i.e.\ the continuous solutions $\Psi(z)$ and
$2/[\Psi(z)+2z]$
(see Table 3.2.2 in [\refBCHI]). Further discretised solutions could in
principle by generated by proceeding up the chain of equations
characterised
by (\eqXXXIX) and (\eqXL) although we shall not do this here.
Nevertheless, what
we have shown is how discrete analogues of complementary error function
solutions of (\eqII) may be obtained.

The solutions arising from equations (\eqXL) are the complementary
error
function-like generalisations of the discrete rational solutions whose
first
few members are listed in (\eqXX). These rational forms vanish as
$z\to\pm\infty$ as do the solutions which are obtained by following the
method
characterised by (\eqXXXVIII)-(\eqXLV). However, we demonstrated in \S2
how
solutions within the so-called `$-2z$' rational hierarchy can be
obtained by
appropriate choice of the separation constant in (\eqXIII); now  we
shall again
take $\mu^{(m)}=-(m+1)\de(2\ga+m\de)$, $m=0,1,\dots$ (with the
corresponding
values of $\alpha$ and $\beta$ then as defined in (\eqasasd)) and the
separation
constant $\kappa_{n+1}^{(m)}=z_n+\ga-m\de$. Then, with the upper choice
of
signs, equations (\eqXIII) show that in this case $P_n^{(m)}$ and
$Q_n^{(m)}$
satisfy
$$\eqalignno{&\left[z_n-(\ga+\de)-m\de\right]P_{n+2}^{(m)}+2(\ga+\de)
P_{n+1}^{(m)}
-\left[z_n+(\ga+\de)+m\de\right]P_n^{(m)}=0,
&\eqnm{eqXLVI}{a}\cr
	     &[z_n-\ga-(m+1)\de]Q_{n+2}^{(m)}+2\ga Q_{n+1}^{(m)}-
[z_n+\ga+(m+1)\de]Q_n^{(m)}=0,&\eqnr{b}\cr}$$
which are identical to (\eqXL) under the transformation $\ga\to -\ga$.
This then provides the starting point for the derivation of further
complementary error function-like solutions which are the natural
extensions of the `$-2z_n$' rational forms whose first few members are
as
in (\eqXXIV). The most straightforward example is provided by $m=0$
whence,
from (\eqXLI), it can shown that
$$Q_n^{(0)}=\thinspace{\rm constant}\thinspace +\thinspace
\left[\Gamma\left({\gamma/\de}\right)\right]^2 \sum_{j=0}^{n-1}
{1\over(\ga+j\de)\Gamma\left(-j+{\ga/\de}\right)\Gamma\left(j+{\ga/\de}
\right)}\eqn{eqXLVII}$$ and we have already shown how this function
asymptotes
to a multiple of $\psi(z)$ in the usual limit. It then follows that
$$(z_n+\ga)P_{n+1}^{(0)}-(z_n+\ga-\de)P_n^{(0)}=2\ga\de Q_n^{(0)}
\eqn{eqXLVIII}$$
and solution of this equation in the limit as $\de\to 0$, with
$\ga\de=1$, leads
to the result that
$$P_n^{(0)}\to P_0^{(0)}-2\de\sum_{j=0}^nQ_j^{(0)}.$$
Thus $P_n^{(0)}$ asymptotes to the function $$P_*^{(0)}=-2\int^z\psi(s)
\thinspace{\d}s= -2z\psi(z)-{\d\psi\over\d z},$$ and so the solution
$$x_n^{(0)}={P_n^{(0)}\over
Q_n^{(0)}}\to-{2z\psi(z)+\psi'(z)\over\psi(z)}
=-2z-\Psi(z)$$ and thus we have the first in the sequence of `$-2z$'
complementary error function solutions of PIV (\eqII). A list of the
simpler
solutions in this hierarchy may be found in Table 3.2.3 of [\refBCHI]
and
discrete equivalents  of these solutions may be derived using the ideas
here;
the essential ingredient of this process is the requirement to solve
equations
(\eqXLVI) for higher values of $m$.

\section{Discussion}
The main result of this paper has been the evaluation of a number of
exact
solutions of the d-PIV (\eqI) and consequently answer a number of the
questions
raised by Tamizhmani \etal [\refTGR]. These solutions are analogues of
known
solutions of the continuous PIV (\eqII). This provides further evidence
of the
close relationship and analogy between properties of the continuous and
discrete
Painlev\'e equations.

We believe that further exact solutions of d-PIV (\eqI)
will arise by studying the discrete \bts\ given in [\refTGR]. A
detailed study
is also required in order to find discrete analogues of these solutions
of PIV
(\eqII) that are expressible in terms of the parabolic cylinder
function
$D_\nu(z)$, for non-integer $\nu$. An important open question for d-PIV
(\eqI)
remains that of the derivation of an associated Lax pair and the form
of the
associated isomonodromic deformation problem.
the

 \ack
PAC is grateful for financial support through U.K.\ Science and
Engineering
Research Council grant GR/H39420. This work was completed whilst APB
was on
leave at the School of Mathematics, University of New South Wales,
Sydney. He is
indebted to the Royal Society and the Australian Research Council
without whose grants his visit would not have been possible. In
addition
he is grateful to the staff and students of New College, UNSW for their
provision of a Visiting Fellowship and to the
School of Mathematics for their hospitality.

\def\refbk#1#2#3#4#5{\item{\hbox to 15pt{[\expandafter \csname
#1\endcsname]\hfill}}{\rm #2},\ {#3} (#4,\ #5).\par}
\def\refjl#1#2#3#4#5#6{\item{\hbox to 15pt{[\expandafter \csname
#1\endcsname]\hfill}}{\rm #2},\ {#3},\ #4\ (#6)\ #5.\par}
\def\refdpp#1#2#3#4{\item{\hbox to 15pt{[\expandafter \csname
#1\endcsname]\hfill}}{\rm #2},\ {#3}\ (#4).\par}
\def\refdppp#1#2#3#4#5{\item{\hbox to 15pt{[\expandafter \csname
#1\endcsname]\hfill}}{\rm #2},\ #4\ #3, {\it Exeter Preprint no.
#5}.\par}
\def\refcf#1#2#3#4#5#6#7{\item{\hbox to 15pt{[\expandafter \csname
#1\endcsname]\hfill}}{\rm #2},\ in: #4, #5, #6 (#3) p.\ #7.}

\references{\frenchspacing
\parindent=20pt
\baselineskip=12pt

\refjl{refRGH}{A. Ramani, B. Grammaticos and J. Hietarinta}
{Phys. Rev. Lett.}{67}{1829}{1991}
\refjl{refRGV}{A. Ramani,\ B. Grammaticos and V. Papageorgiou}
{Phys. Rev. Lett.}{67}{1825}{1991}
\refbk{refInce}{E.L. Ince}{Ordinary Differential Equations}{Dover, New
York}{1956}
\refjl{refASa}{M.J. Ablowitz and H. Segur}{Phys. Rev.
Lett.}{38}{1103}{1977}
\refbk{refAC}{M.J. Ablowitz and P.A. Clarkson}{Solitons, Nonlinear
Evolution Equations and Inverse Scattering, {L.M.S. Lect. Notes Math.},
Vol. {149}}{CUP, Cambridge}{1991}
\refjl{refGromak}{V.I. Gromak}{Diff. Eqns.}{14}{1510}{1977}
\refjl{refBK}{E. Brezin and V. Kazakov}{Phys. Lett. B}{236}{144}{1990}
\refjl{refGM}{D.J. Gross and A.A. Migdal}{Phys. Rev.
Lett.}{64}{127}{1990}
\refjl{refDoug}{M.R. Douglas}{Phys. Lett. B}{238}{176}{1990}
\refjl{refFIK}{A.S. Fokas, A.R. Its and A.V. Kitaev}{Commun. Math.
Phys.}{142}{313}{1991}
\refjl{refPS}{V. Periwal and D. Shewitz}{Phys. Rev.
Lett.}{64}{1326}{1990}
\refjl{refNP}{F.W. Nijhoff and V. Papageorgiou}{Phys. Lett.
A}{153}{337}{1991}
\refjl{refFGR}{A.S. Fokas, B. Grammaticos and A. Ramani}{J. Math.
Anal.
Appl.}{180}{342}{1993}
\refcf{refGR}{B. Grammaticos and A. Ramani}{Kluwer, Dordrecht,
1993}{Applications of Analytic and Geometric Methods to Nonlinear
Differential
Equations}{ed. P.A. Clarkson}{{NATO ASI Series C: Mathematical and
Physical Sciences\/}, Vol. {413}}{299}
\refjl{refKOSGR}{K. Kajiwara, Y. Ohta, J. Satsuma, B. Grammaticos and
A.
Ramani}{J. Phys. A: Math. Gen.}{27}{915}{1994}
\refjl{refGNPRS}{B. Grammaticos, F.W. Nijhoff, V. Papageorgiou, A.
Ramani and J. Satsuma}{Phys. Lett. A}{185}{446}{1994}
\refjl{refPNGR}{V. Papageorgiou, F.W. Nijhoff, B. Grammaticos and A.
Ramani}{Phys. Lett. A}{164}{1825}{1992}
\refdpp{refBCHI}{A.P. Bassom,\ P.A. Clarkson\ and A.C.
Hicks}{``B\"acklund
transformations and solution hierarchies for the fourth \p\ equation'',
Stud.
Appl. Math., to appear}{1994}
\refdpp{refBCHII}{A.P. Bassom,\ P.A. Clarkson\ and A.C. Hicks}
{``On the application of solutions of the fourth Painlev\'e equation to
various
physically motivated nonlinear partial differential
equations'', preprint {\bf M94/32}, Department of Mathematics,
University of
Exeter}{1994}
\refjl{refMurata}{Y. Murata}{Funkcial. Ekvac.}{28}{1}{1985}
\refjl{refTGR}{K.M. Tamizhmani,\ B. Grammaticos and A. Ramani}
{Lett. Math. Phys.}{29}{49}{1993}
\refbk{refASte}{M. Abramowitz and I.A. Stegun}{Handbook of Mathematical
Functions}{Dover, New York}{1965}
\refjl{refLukash}{N.A. Lukashevich}{Diff. Eqns.}{3}{395}{1967}
 \refjl{refBCHM}{A.P. Bassom, P.A. Clarkson, A.C. Hicks and J.B.
 McLeod}{Proc. R.
Soc. Lond. A}{437}{1}{1992}
 \refjl{refBCH}{A.P. Bassom, P.A. Clarkson and A.C. Hicks}{IMA J.
 Appl.
Math.}{50}{167}{1993}

}

\bye